\documentstyle[prl,aps,graphicx,twocolumn]{revtex}

\newcommand{\myscalebox}[1]{\scalebox{0.34}[0.45]{#1}}
\begin{document}
\draft
\title{Evidence for existence of many pure ground states in
 3d $\pm J$ Spin Glasses}

\author{Alexander K. Hartmann}

\address{{hartmann@tphys.uni-heidelberg.de}\\
{Institut f\"ur theoretische Physik, Philosophenweg 19, }\\
{69120 Heidelberg, Germany}\\
{Tel. +49-6221-549449, Fax. +49-6221-549331}}

\date{\today}
\maketitle

\begin{abstract}

Ground states of 3d EA Ising spin glasses are calculated
for sizes up to $14^3$ using a combination 
of genetic algorithms and cluster-exact
approximation . The distribution $P(|q|)$ of overlaps is calculated.
For increasing size the width of $P(|q|)$ converges
to a nonzero value, indicating that many pure ground states exist
for short range Ising spin glasses.

\end{abstract}


\paragraph*{Introduction}

The behavior of the Edwards-Anderson (EA) $\pm J$ Ising spin glass with short
range (i.e. realistic) interactions is still not well understood.
The EA Ising spin glass is a system of $N$ spins 
$\sigma_i = \pm 1$, described by the Hamiltonian

\begin{equation}
H \equiv - \sum_{\langle i,j\rangle} J_{ij} \sigma_i \sigma_j
\end{equation}

The sum $\langle i,j\rangle$ goes over nearest neighbors. In this letter we
consider 3d cubic systems with periodic boundary conditions,
$N=L^3$ spins and the exchange interactions (bonds) take
$J_{ij} = \pm 1$ with equal probability under the constraint
$\sum_{<i,j>} J_{ij} = 0$.

One of the most important questions is, whether many pure states exist
for realistic spin glasses. For the infinite ranged Sherrington-Kirkpatrik
(SK) Ising spin glass this question was answered positively by the
replica-symmetry-breaking mean-field (MF) scheme by Parisi \cite{parisi2}.
Some authors \cite{newman} disbelieve that the MF theory is valid in total
for realistic spin glasses. But also a complete different model is proposed:
the Droplet Scaling (DS) theory 
\cite{mcmillan,bray,fisher1,fisher2,bovier}
suggests that only two pure states (related by a global flip) exist and
that the most relevant excitations are obtained by reversing large
domains of spins (the droplets).
In \cite{berg1} ground states were calculated using multicanonical sampling,
but no decision between MF and DS could be made. Recent results of
Monte Carlo (MC) simulations \cite{marinari} at finite 
temperature find evidence for the MF picture.

In this letter we show our results for the calculation of spin glass
ground states using a hybrid of genetic algorithms \cite{pal1,michal} 
and cluster-exact
approximation (CEA) \cite{alex2}. CEA is a discrete optimization
procedure which is based on the construction of spin clusters exhibiting
no frustration and calculating exact ground states for these clusters
using graph theoretical methods. By calculating the distribution of
overlaps for different sizes $L$ we find
evidence for the existence of many pure ground states for the
$\pm J$ spin glass.

\paragraph*{Observables}

For a fixed realization $J=\{J_{ij}\}$ of the exchange interactions and two
statistical independent configurations (replicas)
$\{\sigma^{\alpha}_i\}, \{\sigma^{\beta}_i\}$, the overlap \cite{parisi2}
is defined as

\begin{equation}
q^{\alpha\beta} \equiv \frac{1}{N} \sum_i \sigma^{\alpha}_i \sigma^{\beta}_i 
\end{equation}

The ground state of a given realization is characterized by the probability
density $P_J(q)$. Averaging over the realizations $J$, denoted
by $[\,\cdot\,]_{av}$, results in ($Z$ = number of realizations)

\begin{equation}
P(q) \equiv [P_J(q)]_{av} = \frac{1}{Z} \sum_{J} P_J(q)
\end{equation}

Because no external field is present the densities are symmetric:
$P_J(q) = P_J(-q)$ and $P(q) = P(-q)$. So only averages of $|q|^n$ are
relevant:

\begin{eqnarray}
\overline{|q_J|^n} & \equiv & \int_{-1}^1 |q|^n P_J(q) dq \label{def_mean_q}\\
\overline{|q|^n} & \equiv & \int_{-1}^1 |q|^n P(q) dq 
\end{eqnarray}

The Droplet model predicts that only two pure states exist, implying
that $P(q)$ converges for $L \to \infty$ to 
$P(q) = \frac{1}{2}(\delta(q-\overline{|q|})+\delta(q+\overline{|q|}))$
(we don't indicate the $L$ dependence by an index), while in the MF picture
the density remains nonzero for a range $-q_1 \le q \le q_1$ with
peaks at $\pm q_{\max}$ ($0< q_{\max} \le q_1$). Consequently the variance

\begin{equation}
\sigma^2(|q|) \equiv \int_{-1}^1 (\overline{|q|} - |q|)^2P(q)dq = 
\overline{|q|^2} - \overline{|q|}^2 \label{def_sigma_q}
\end{equation}

stays finite for $L \to \infty$ in the MF pictures 
($\sigma^2(|q|) = a_1+a_2/L^3$) while 
$\sigma^2(|q|) \sim L^{-y} \to 0$ according the DS approach. Here $y$ is the
zero-temperature scaling exponent \cite{mcmillan} denoted $\Theta$ in 
\cite{fisher1,fisher2}.

We have measured the energy per spin $e$ as well. 
 Additionally to the usual fits other scaling corrections were
discussed \cite{pal1}, including an exponential one 
($e = a_{\infty} + a_1 \exp(-a_2 L)$).

\paragraph*{Results}

We performed ground state calculations for sizes $L=3,4,5,6,8,10,12,14$. 
We calculated from 100 realizations for $L=14$ up to 4000 realizations
for $L=3$. One $L=14$ run needs typically 540 CPU-min on a 80MHz PPC601
processor
(70 CPU-min for $L=12$, $\ldots$, 0.2 CPU-sec for $L=3$).
For each
realization $40$ independent runs were made. Each run resulted in one 
configuration, which was
stored, if it exhibited the ground state energy. For $L=14$ this resulted
in an average of $9.9$ ground states per realization while for $L=3$ we 
got an average of $39.9$ ground states. Details of the algorithm, simulation
parameters and more results are given in \cite{alex3}.

The results of the calculations are summarized in table \ref{tab_results}.

Because we want to estimate the quality of the minimization
procedure, we discuss
the average ground state energy at first. By comparison with
previous calculations \cite{berg1,pal1} we can see that we have obtained
the same or slightly lower (i.e. better) energy, except for $L=12$ in
\cite{berg1} which is based only on 7 realizations. So we are confident that
we get very close to the true ground states using our approximation procedure.

We extrapolate to $L\to\infty$ by using finite size scaling (FSS) fits. 
We used the Levenberg-Marquard Method implemented in \cite{numrec}
which gives also the probability Q of finding such a bad data-set (provided
that the fit function is exact). 

Table \ref{tab_results3} show the results for FFS fits
for the following three functions:

\begin{eqnarray}
f(L) & \equiv & a_{\infty} + a_1/L^3 \\
g(L) & \equiv & a_{\infty} + a_1L^{-a_2} \\
h(L) & \equiv & a_{\infty} + a_1 \exp(-a_2 L)
\end{eqnarray}

For $\sigma^2(|q|)$ results for fits of
the first two functions $f(L)$ and $g(L)$ (with $a_{\infty}$ set to 0) are
given in the same table as well.

We get a ground state energy of the infinite system of 
$e_{\infty} =a_{\infty}= -1.7868(3)$ (Q=0.20) resp. 
$e_{\infty} = a_{\infty} = -1.7876(3)$
with $a_2=2.84(3)$ (Q=0.77). The exponential fit is ruled out (Q=$10^{-7}$).
Our ground state energy is lower than some previous results:
$e_{\infty} = -1.75$ \cite{kirkpatrick}, 
$e_{\infty} = -1.76(2)$ \cite{morgenstern}, and of the same magnitude as the
most accurate study so far
$e_{\infty} = -1.7863(4)$ \cite{pal1}. In \cite{berg1} a lower value
was found ($e_{\infty} = -1.839(4)$), but that study suffers from only
seven $L=12$ realizations.

\begin{figure}[ht]
\begin{center}
\myscalebox{\rotatebox{-90}{\includegraphics{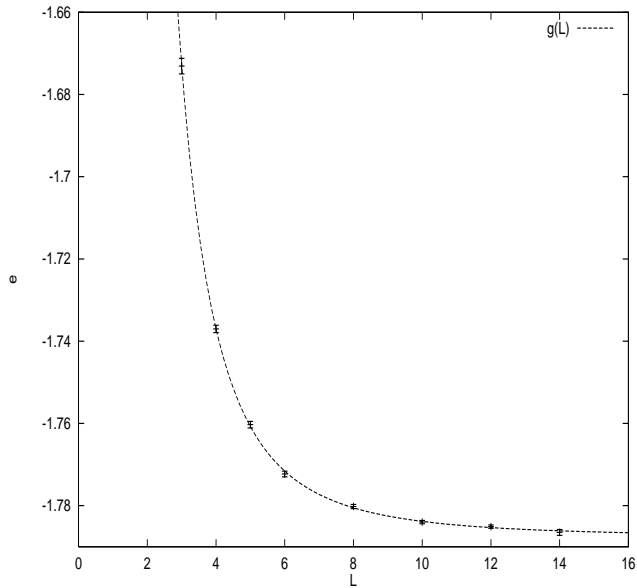}}}
\end{center}
\caption{Ground state energy for $3\le L \le 14$ and best FSS fit}
\label{fig_e_min}
\end{figure}

The result for the energy, together with the best fit,
is displayed in fig. \ref{fig_e_min}.

In \cite{pal1} a $L^{-3}$ fit was ruled out, instead a $L^{-2.97}$ fit
was preferred. Our data shows a similar tendency, but the data for the
energy is not accurate enough to decide this question. The second fit
supports the MF picture too, because the value $y=a_2=2.84$ of the exponent
exceeds the limit $y\le (d-1)/2$ proposed by DS.
Since the aim of this letter is to distinguish whether only two
or infinitely many  pure ground states exist, we proceed by
analyzing the order parameter distribution $P(|q|)$. 

\begin{figure}[ht]
\begin{center}
\myscalebox{\rotatebox{-90}{\includegraphics{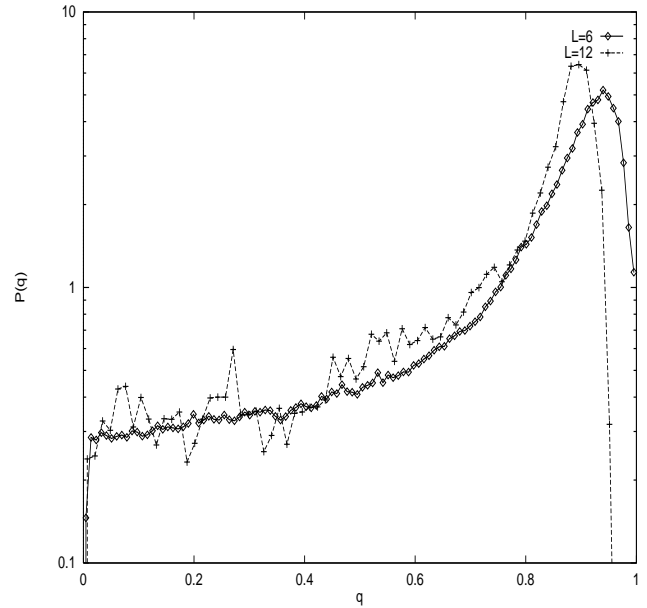}}}
\end{center}
\caption{$P(|q|)$ for $L=6,12$}
\label{fig_plq}
\end{figure}

\begin{figure}[ht]
\begin{center}
\myscalebox{\rotatebox{-90}{\includegraphics{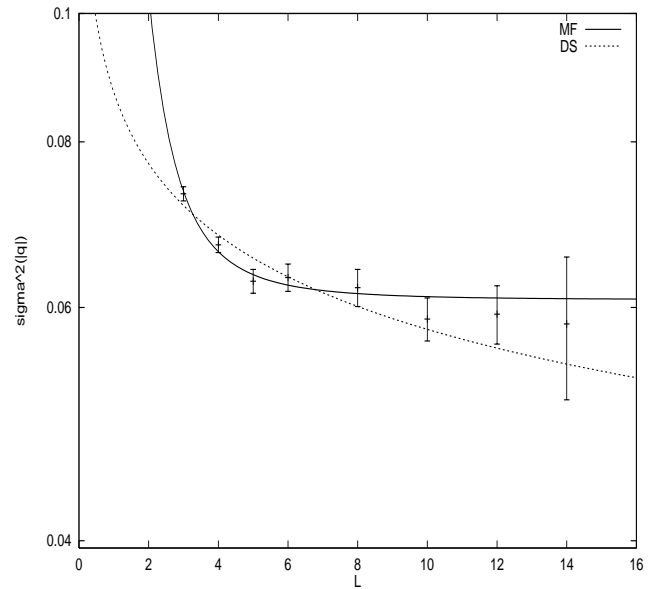}}}
\end{center}
\caption{Variances $\sigma^2(|q|)$
for $3\le L \le 14$ and FSS fits}
\label{fig_sigma}
\end{figure}

In
fig.\ \ref{fig_plq} the probability density of the overlap is
displayed for $L=6$ and $L=12$. Each realization enters the
sum with the same weight, even if different numbers of ground states
were available for the calculation of $P_J(|q|)$.
Only a small difference for high $q$-values is visible, which is reflected
also in a slight shift of the average to smaller values 
(see table \ref{tab_results}). But especially for $|q|\le 0.8$ no significant
reduction in the probability is visible.
So many different ground states still exist
for higher $L$. This result is
supported quantitatively by the variances of $|q|$ as function of $L$ which are
shown in fig.\ \ref{fig_sigma}. For all sizes the magnitude of the values is
higher than in \cite{berg1}, indicating that their MC algorithm
does not explore the configuration space sufficiently.
Our fits show a higher probability
for the existence of many pure states ($Q=0.79$) against the two state
picture ($Q=0.09$). 

Another interesting questions is whether the difference between the 
density functions of single realizations and the average $P(|q|)$ remains
nonzero
for increasing $L$, i.e. whether the short range spin glasses exhibit
lack of self-averaging. But presently we have not computed 
enough ground states per realization for large sizes to address this question.

\paragraph*{Conclusion}

By the calculation of ground states using genetic cluster-exact approximation
we find evidence for the existence of many pure states in short range
spin glasses. The major drawback of our method is that the ground states
are not exact, but for these lattice sizes this is beyond the power of
actual algorithms and computers. By comparison with other results 
we feel confident that the energy of our states is very close to the 
true ground state energy. 
In the moment we do not have enough ground states per
realization calculated (for higher $L$)
to decide whether short range spin glasses exhibit lack of
self-averaging or not.

\paragraph*{Acknowledgements}

We would like to thank D.W. Heermann for manifold support and
 G. Reinelt for fruitful discussions. We thank G. P\"atzold for
critical reading of the manuscript and
the {\em Paderborn Center for Parallel Computing}
 for the allocation of computer time. This work was supported
by the Graduiertenkolleg ``Modellierung und Wissenschaftliches Rechnen in 
Mathematik und Naturwissenschaften'' at the
{\em In\-ter\-diszi\-pli\-n\"a\-res Zentrum f\"ur Wissenschaftliches Rechnen}
 in Heidelberg.

\newpage

\begin{table}[ht]
\begin{center}
\begin{tabular}{cddd}
\hline
$L\rule[-2mm]{0mm}{0.7cm}$ & $e$ & $\overline{|q|}$ &
$\sigma^2(|q|)$  \\ \hline
 3 & -1.6731(19) & 0.742( 3) & 0.0731( 9) \\
 4 & -1.7370( 9) & 0.748( 3) & 0.0669( 9) \\
 5 & -1.7603( 8) & 0.743( 4) & 0.0628(13) \\
 6 & -1.7723( 7) & 0.742( 4) & 0.0632(15) \\
 8 & -1.7802( 5) & 0.726( 6) & 0.0621(20) \\
10 & -1.7840( 4) & 0.737( 6) & 0.0588(22) \\
12 & -1.7851( 4) & 0.698( 8) & 0.0593(30) \\
14 & -1.7865( 7) & 0.701(20) & 0.0583(72) \\
\end{tabular}
\end{center}
\caption{Results.}
\label{tab_results}
\end{table}

\begin{table}[ht]
\begin{center}
\begin{tabular}{cccccc}
\hline
\small
 & fit & $a_{\infty}$ & $a_1$ & $a_2$ & Q \\ \hline
$e$ & f(L) & -1.7869(3) & 3.15( 5) & -       & 0.20 \\
$e$ & g(L) & -1.7876(3) & 2.60(15) & 2.84(5) & 0.77 \\
$e$ & h(L) & -1.7846(8) & 1.01(17) & 0.75(5) & $10^{-7}$ \\
$\sigma^2(|q|)$ & f(L) & 0.0608(6) & 0.34(3) & - & 0.79 \\
$\sigma^2(|q|)$ & g(L) & 0         & 0.088(4)& 0.18(3) & 0.09 \\
\end{tabular}
\end{center}
\caption{Result of FSS-fits.}
\label{tab_results3}
\end{table}

\end{document}